\documentclass{llncs}
\usepackage{makeidx}  
\usepackage{graphicx}
\usepackage{amsmath}
\usepackage{amssymb}
\usepackage[colorinlistoftodos]{todonotes} 
\usepackage{stmaryrd}
\usepackage{tabularx}
\begin{document}
\frontmatter          
\pagestyle{headings}  
\addtocmark{Hamiltonian Mechanics} 
\mainmatter              
\title{Deep Boosted Regression for MR to CT Synthesis}

\author{Kerstin Kl\"{a}ser$^1$ \and Pawel Markiewicz$^1$ \and Marta Ranzini$^1$ \and Wenqi Li$^2$ \and Marc Modat$^{2,1}$ \and Brian F Hutton$^3$ \and David Atkinson$^4$ \and Kris Thielemans$^3$ \and M Jorge Cardoso$^{2,1}$ \and S\'{e}bastien Ourselin$^2$ }
\institute{$^1$Centre for Medical Image Computing, University College London, London, UK \\ $^2$School of Biomedical Engineering and Imaging Sciences, King's College London \\ $^3$Institute of Nuclear Medicine, University College London, London, UK \\ $^4$Centre for Medical Imaging, University College London, London, UK}

\maketitle

%
%
\begin{abstract}
Attenuation correction is an essential requirement of positron emission tomography (PET) image reconstruction to allow for accurate quantification. However, attenuation correction is particularly challenging for PET-MRI as neither PET nor magnetic resonance imaging (MRI) can directly image tissue attenuation properties. MRI-based computed tomography (CT) synthesis has been proposed as an alternative to physics based and segmentation-based approaches that assign a population-based tissue density value in order to generate an attenuation map. We propose a novel deep fully convolutional neural network that generates synthetic CTs in a recursive manner by gradually reducing the residuals of the previous network, increasing the overall accuracy and generalisability, while keeping the number of trainable parameters within reasonable limits. The model is trained on a database of 20 pre-acquired MRI/CT pairs and a four-fold random bootstrapped validation with a 80:20 split is performed. Quantitative results show that the proposed framework outperforms a state-of-the-art atlas-based approach decreasing the Mean Absolute Error (MAE) from 131HU to 68HU for the synthetic CTs and reducing the PET reconstruction error from 14.3\% to 7.2\%.
\end{abstract}
\section{Introduction}

Positron emission tomography - magnetic resonance imaging (PET-MRI) is a relatively new joint imaging technique that combines the functional information from PET with the flexibility of MRI. To obtain quantitative PET images, it is essential to know the tissue attenuation coefficients throughout the patient. However, this is a difficult problem for PET-MRI as neither PET nor MRI can directly image tissue attenuation properties, which is why computer tomography (CT) remains the clinically accepted gold-standard for attenuation correction. However, it is desirable to circumvent the requirement of an additional CT acquisition not just to reduce the exposed radiation dose to the patient but also to avoid the risk of registration errors between the MR and CT volumes. Hence, synthesising pseudo CTs from MR images gained a lot of interest in the field of attenuation correction for hybrid PET-MR systems. 

Within the last ten years, several research groups focused on the development of single- and multi-atlas-based approaches that predict attenuation coefficients on a continuous scale by deforming an anatomical model or dataset to match the subject's anatomy using non-rigid registration. Synthesis methods based on multi-atlas information propagation, such as the model proposed by Burgos et al. \cite{Ninon}, have dominated this area of research for several years. 

Recently, deep learning approaches have started outperforming multi-atlas methods. In particular, convolutional neural networks (CNNs) have proved to be a powerful tool for translating an image between domains (as between MRI and CT). Within deep learning approaches, methods for image-to-image translation can be classified into two classes: unsupervised and supervised representation learning. 
The first learns the contextual information between two image domains from unpaired data, allowing general-purpose image-to-image translation. For example, Zhu et al. recently proposed a CycleGAN model that assumes an underlying relationship between two different domains that can be learned by an adversarial loss that competes with a second network trained to produce images that are in principle indistinguishable from the desired output \cite{CycleGAN}. Recently, Wolternik et al. successfully applied the CycleGAN model to medical image data in order to perform CT synthesis \cite{Wolternik}. However, just like the majority of CNN frameworks, their framework addresses the image translation problem on the basis of 2D image representations, neglecting the 3 dimensional nature of the anatomical representation. Several attempts have been made to stably train 3 dimensional networks, a challenging task due to the curse of dimensionality. Most 3D network architectures exploit a fully convolutional architecture, where neighbourhood information is preserved either through pooling/upsampling layers \cite{Net1,Net2}, or through the use of dilated convolutions \cite{Wenqi}.
Here, we approach the image translation task in a supervised learning setting, where corresponding data pairs are available. However, unlike previous supervised methods \cite{XiaoHan}, we propose to use a fully 3D architecture with an efficient parameter count and large receptive field, namely HighRes3DNet by Li et al. \cite{Wenqi}, to learn a 3 dimensional representation of the data. This representation is then mapped to the domain of CT images through a series of 1D convolutions with non-linear activation functions. This proposed architecture also makes extensive use of residual connections to avoid the need to model the identity mapping of the representation, improving the overall accuracy and training stability. Finally, we reformulate the residual connection architecture as a corrective model, which can be seen as a form of boosting in classic machine learning. This is achieved by recursively applying a corrective model with shared parameters and with a deep supervision loss, recursively reducing the residuals of the predictions. 
We evaluate our approach on a dataset of 20 patients using a four-fold random bootstrapped validation with a 80:20 split. The results demonstrate an improvement over a state-of-the-art multi-atlas based method, as well as the ability of our method to simulate abnormal structures not observable in the training data. As we are validating the advantages of the use of a recursive boosting model, the contribution of the paper is independent of the choice of cost function. 

\section{Methods}

\subsection{Deep Boosted Regression}
The aim of the proposed image synthesis approach is to find a mapping from the domain of T1- and T2-weighted MR input images to the domain of CT images. This mapping can be formulated as 

\[ 
\mathbb{R} ^{T_1,T_2} \to \mathbb{R}^{CT},
\]
which is a mapping from $y \mapsfrom f(x)$, where $f$ is a function that maps input $x \in \mathbb{R} ^{T_1,T_2} $ to $y \in \mathbb{R} ^{CT}$. This mapping function is highly nonlinear, and can be approximated by a composition of simpler functions with parameters $\phi$, of the form $\tilde{y} = f^{(n)}(f^{(n-1)}(...(f^{(2)}(f^{(1)}(x, \phi_1), \phi_2),... ),\phi_{n-1}),\phi_n)$. In a supervised learning context, these parameters $\phi$ are determined by minimising a loss function that aims to minimise the residuals between the predicted CT $\tilde{y}$ and the true CT $y$

\[ 
\mathcal{L}_2 = ||y - \tilde{y} ||_2 .
\]

Note, however, that the large number of functions and parameters $\phi$ creates computational and optimisation challenges. To avoid this, we propose to formulate the problem as a boosting model, where the output of each function $f^{(n)}$ aims to approximate $y$. If $\tilde{y}_1=f^{(1)}(x, \phi_1)$, then subsequent functions can be seen as a form of corrective learning, as $\tilde{y}_2=f^{(2)}(\tilde{y}_1, x)$. Thus, the model above can be rewritten as
\[ 
\tilde{y} = f^{(n)}(f^{(n-1)}(...(f^{(2)}(f^{(1)}(x, \phi_1), x, \phi_2),... ), x,\phi_{n-1}), x,\phi_n).
\]
It is important to note that this corrective learning model introduces more parameters for every corrective function $f$, resulting in model overfit and making it hard to optimise. Instead, we propose to create a single corrective function $f^{(c)}$, equivalent to sharing parameters between functions $f^{(2)}$ to $f^{(n)}$, which is applied recursively after an initial approximation of $\tilde{y}$ given by $f^{(1)}$. 
We can define our recursion as  
\[ 
\tilde{y}_k = 
	\begin{cases} 
      f^{(1)}(x \; | \; N_1) & \; if \; k =0 \\
     
      f^{(c)}(x, \; \tilde{y}_{k-1} \; | \; N_c) & \; if \; k >0
   \end{cases}
\]
where a function with parameters $N_1$ synthesises $\tilde{y}_1$ from an input MRI $x$, at iteration $k=0$. For $k>0$, a corrective function, with parameters $N_c$, maps the previous prediction $\tilde{y}_{k-1}$ and the input MR images $x$ to a better approximation of the true CT $y$. 
Finally, to ensure that the function's parameters can be optimised, we change the loss function to

\newcommand\norm[1]{\left\lVert#1\right\rVert}
\[ 
Loss = \sum_{k=0}^{n}\norm{\tilde{y}_k-y}^2  .
\]
thus providing a form of deep supervision by introducing gradients for each function $f$. We called this method Deep Boosted Regression as it is inspired by the recursive residual minimisation approach of classical boosting models.

\subsection{Proposed Network Architecture}

The functions described in the previous section are approximated by two separate CNNs, both following the network architecture of the high-resolution compact architecture presented by Li et al. \cite{Wenqi}, which has been shown to be very efficient in learning 3D representations from large-scale image data. It consists of 20 convolutional layers with kernel size 3 x 3 x 3 that encode low-level image features. Mid- and high-level image features are captured within the following convolutional layers with kernels that are dilated by a factor of two or four, respectively, preserving the spatial resolution of the input image throughout the network. Convolutional layers are grouped into pairs of two, and residual connections are added that enable an identity mapping so that both parameters and computational cost are minimal as shown by He et al. \cite{He}. 

\begin{figure}[t!]
\begin{center}
\includegraphics[width=0.7\textwidth]{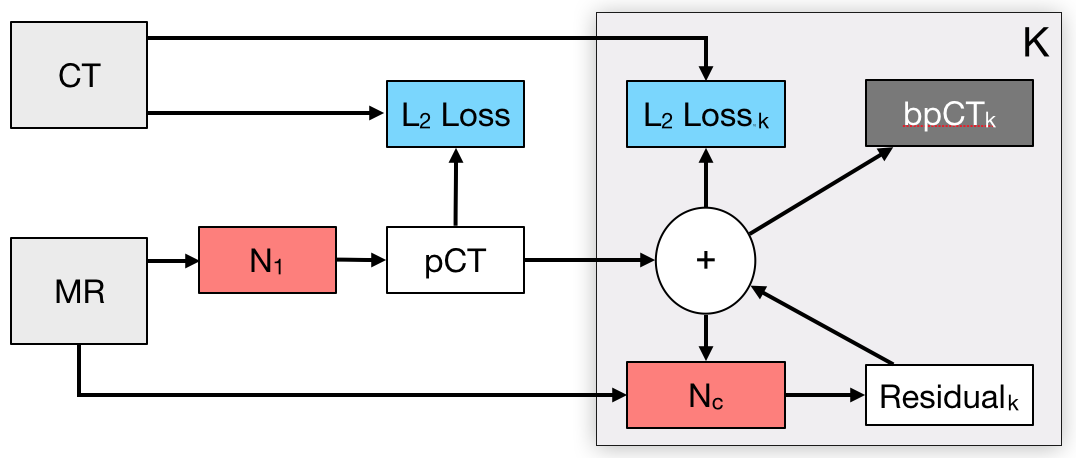}
\caption{Framework of proposed Deep Boosted Regression method. MRs are fed into a first network $N_1$, an initial pseudo CT (pCT) is synthesised by minimising the loss between pCT and original CT. Within the space K, residual learning is performed, where the residuals are added to pCT and fed into a second network $N_c$, wherefore the "+" illustrates an accumulator. A second loss is introduced minimising the difference between ground-truth CT and updated pCT. The final output is an error boosted pCT (bpCT). The number of residual learning cycles (K) is limited to avoid overfitting (e.g. we used K=4).}
\label{architecture}
\end{center}
\end{figure}

The proposed network architecture is illustrated in Fig. \ref{architecture}. The first network $N_1$ is trained to synthesise an initial pseudo CT (pCT) taking both T1- and T2-weighted MR images as inputs. This first pCT is passed to a second network $N_c$ that learns the residuals between pCT and the real CT. Therefore the weights of $N_c$ depend on the output of $N_1$, but not vice versa. An improved pCT is then generated by adding the residuals to the initially synthesised pCT, which is then again fed back into $N_c$ in order to update the weights of the network. By sharing the parameters of $N_c$ no additional parameters are introduced to the network keeping computational complexity within limits and making the model more generalisable even if only a limited number of training datasets are available. This recursive cycle can be repeated for $k$ iterations, however, the number of iterations is limited to avoid overfitting. The proposed Deep Boosted Regression (DBR) approach exploits the advantages of the recursive boosting model and is therefore independent of the choice of the cost function.

\subsection{Implementation Details}

\begin{figure}[b!]
\begin{center}
\includegraphics[width=\textwidth]{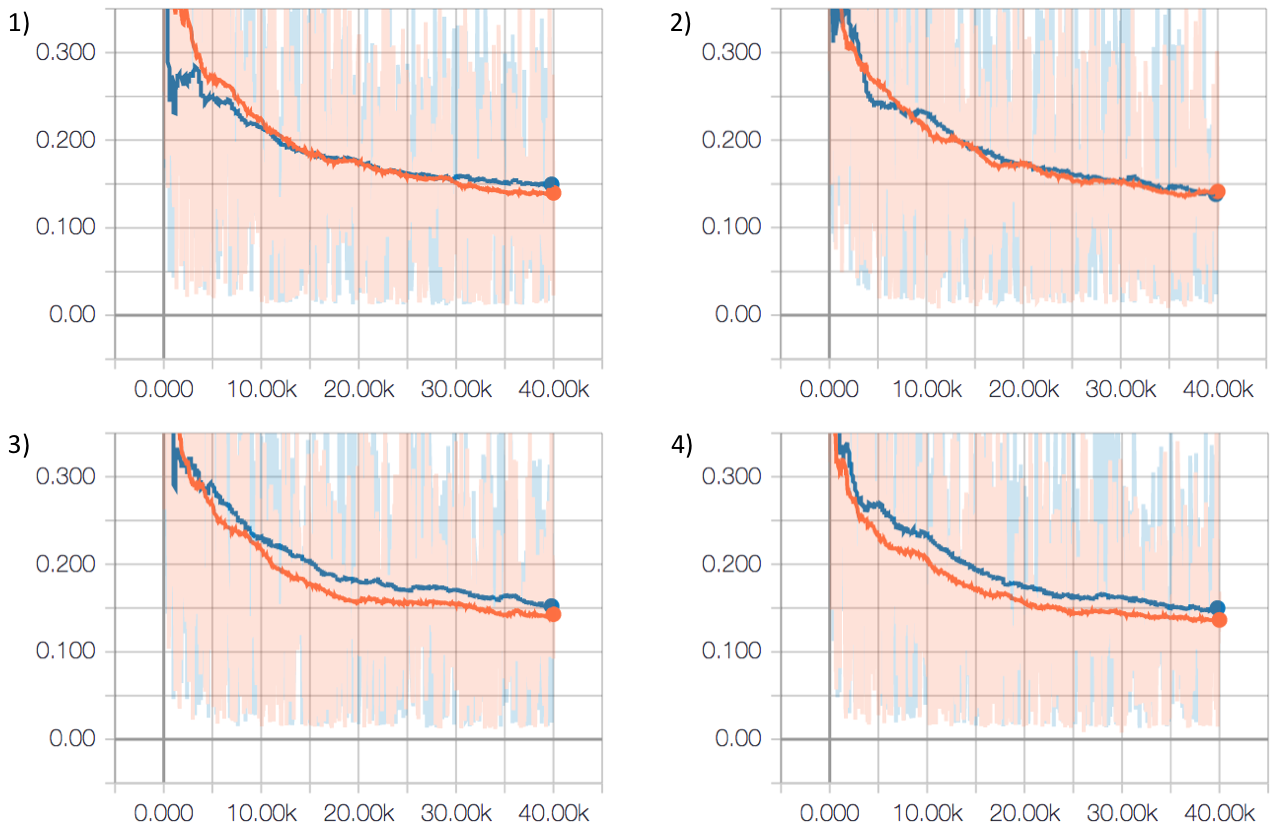}
\caption{Training (orange) and validation (blue) loss for each fold of four-fold random bootstrapped validation with a 80:20 split. The training was terminated after 40K iterations when the error had converged.}
\label{LearningCurves}
\end{center}
\end{figure}

In the training stage, the data (see sec. \ref{section3}) were randomly sampled into subvolumes of size 56 x 56 x 56 pixels that were augmented by randomly rotating each of the three orthogonal planes on the fly by an angle in the interval of [-10$^{\circ}$, 10$^{\circ}$]. The MR data was also randomly scaled by a factor between 0.9 and 1.1. Patches were sampled more often from high frequency regions of the image as these areas are harder to model. The network was trained from scratch on a single NVIDIA Titan X GPU using the Adam optimisation method. We did a four-fold random bootstrapped validation, where for each fold, the data was split into 70$\%$ training, 10$\%$ validation and 20$\%$ testing data. The model was trained with a learning rate of 0.001 and a weight decay of 5.0 x 10$^{-8}$. We trained the network for 10K iterations before decreasing the learning rate by a factor of 10. We terminated the training after 40K iterations when the error had converged. The training and validation loss is demonstrated in Fig. \ref{LearningCurves}. We implemented our method with NiftyNet, which is a TensorFlow-based open-source CNN platform that can be used for research in medical image analysis. The model will be made available online as part of the NiftyNet model zoo \cite{NiftyNet}.

\section{Experimental Datasets and Materials}
\label{section3}
The experimental dataset consisted of pairs of T1- and T2-weighted MR and CT brain images of 20 patients. For each subject MRs and CTs were aligned using first a rigid registration algorithm followed by a very low degree of freedom non-rigid deformation \cite{Ninon}. A second non-linear registration was performed, using a cubic B-spline with normalised mutual information, only on the neck region to correct for soft tissue shift \cite{Modat}. Each volume had 301 x 301 x 153 voxels with a voxel size of approximately 1$mm^3$. For evaluation purposes a head region mask was extracted from the CT image to exclude the background from the analysis.

\section{Experiments and Results}

\begin{figure}[t!]
\includegraphics[width=\textwidth]{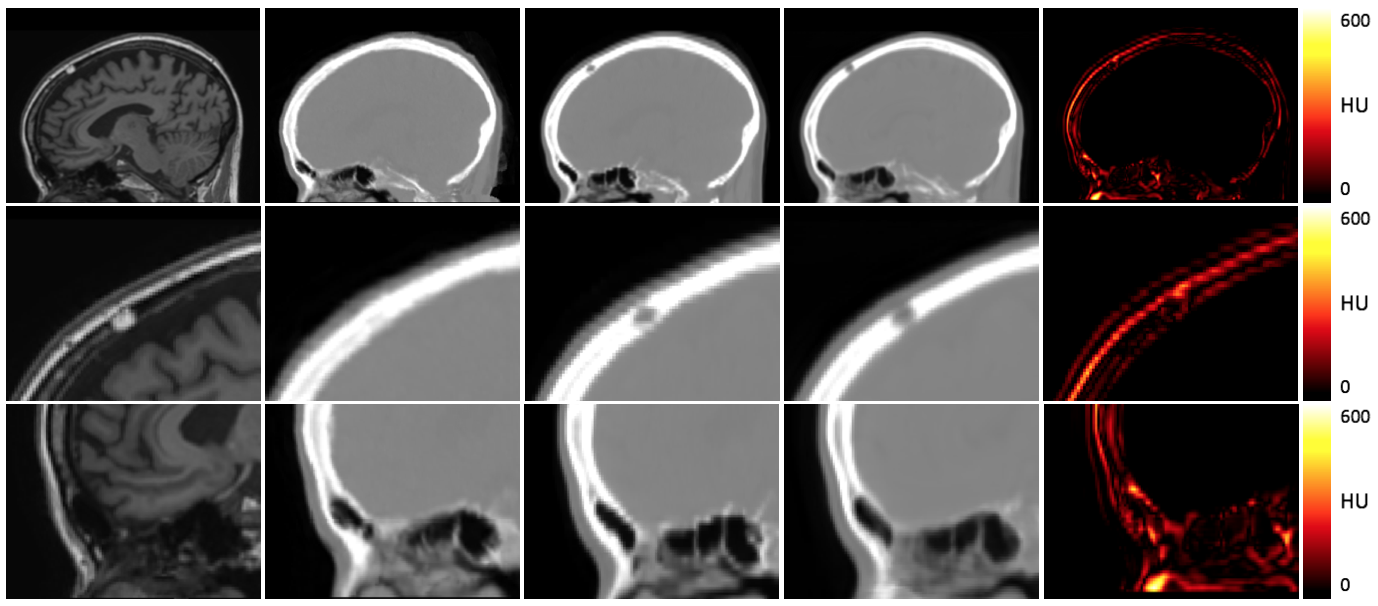}
\caption{\textit{From left to right:} Input MR image, synthesised CT using multi-atlas propagation approach, reference real CT, synthesised CT using proposed Deep Boosted Regression, and absolute error between real and synthesised boosted CT images of the whole head (top), an anatomical abnormality in the skull (middle) and the sinus region (bottom).}
\label{results_1}
\end{figure}

Figure \ref{results_1} shows an example MR input image, a synthesised CT image obtained by a current state-of-the-art multi-atlas propagation approach \cite{Ninon}, a synthesised CT generated by the proposed deep boosting approach and the corresponding reference CT images. Other than the multi-atlas propagation method, our network is able to generate details in the pseudo CT that the network has never seen. For example, Fig. \ref{results_1} shows a patient with an epidermoid cyst in the skull being correctly generated by the network, even though no other patient in the training database shows a similar anatomical abnormality. The greatest error can be observed at the contour of the head and air, especially in the region of the nasal cavity. 

\begin{figure}[b!]
\includegraphics[width=\textwidth]{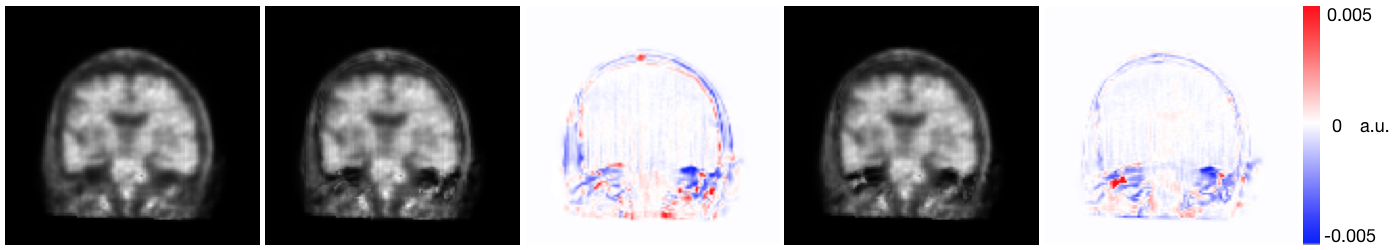}
\caption{\textit{From left to right:} PET reconstructed with real CT, with synthesised CT using multi-atlas propagation approach (mapCT), difference between real CT and mapCT, PET reconstructed with synthesised CT from Deep Boosted Regression (bpCT), difference between real CT and bpCT.}
\label{pet}
\end{figure}

We had no access to the raw PET data therefore the PET images were reconstructed using the following simulation using NiftyPET software \cite{Pawel}. The original PET image was forward projected using the Siemens mMR scanner geometry, then multiplied by the forward projected CT-based attenuation map in order to obtain simulated measured PET sinograms. The simulated measured data were then reconstructed using the original CT-based attenuation map to obtain a reference image, to which the reconstructed images obtained by the multi-atlas propagation method and our Deep Boosted Regression approach were compared. Figure \ref{pet} shows an example slice of the simulated reference PET, the synthesised PET and the corresponding difference image generated by the multi-atlas propagation method and the proposed DBR approach, respectively. 

\begin{figure}[t!]
\centering
\includegraphics[width=\textwidth]{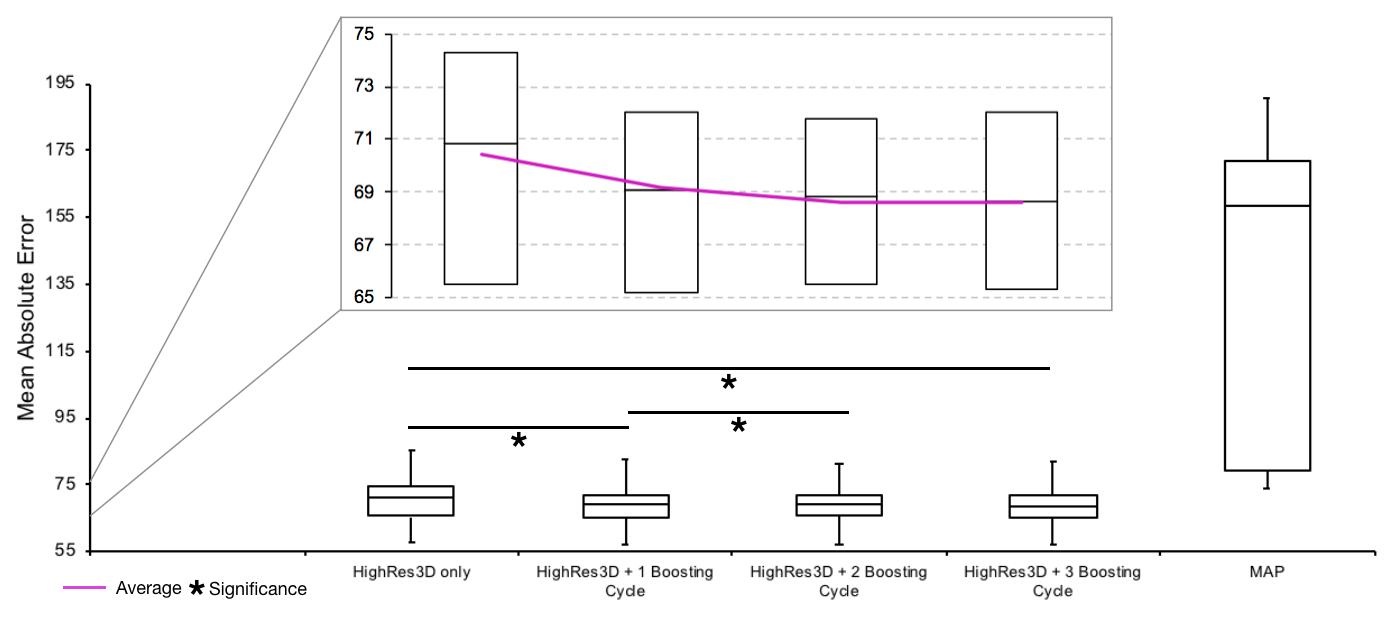}
\caption{Progression of Mean Absolute Error (MAE) of synthesised CTs after each step of the Deep Boosted Regression network (zoomed panel) compared to a current state-of-the-art multi-atlas propagation method (MAP). The MAE decreases significantly after the first and second boosting cycle (horizontal lines with asterisk) as well as overall compared to a simple feed forward network (HighRes3D only).}
\label{box}
\end{figure}

In order to quantify the results we calculated the Mean Absolute Error (MAE) of the synthesised CTs only within the head region by masking the surrounding air out and compared it to the multi-atlas propagation method. The choice of MAE as error metric derives from its good suitability for PET attenuation correction and due to the quantitative nature of CT images. We also investigated how the MAE of the testing data progresses after each run through the network. The results are demonstrated in Fig. \ref{box}. The average MAE of the test images synthesised with the multi-atlas propagation approach lies around 131.4HU, whereas the proposed method for MR-to-CT translation is able to reduce this error by around 48\%. A paired t-test was used to show that the agreement between true CT images and images generated by the proposed model was significantly higher ($p<10^{-5}$) compared to the images synthesised using the multi-atlas propagation approach. The MAE also significantly reduces after the first two boosting cycles of the network confirming that the integrated boosting for the minimisation of the residuals works. Table \ref{MAE} shows a direct comparison between the proposed model, the multi-atlas propagation approach and two other recent deep learning methods for MR to CT synthesis.

\begin{table}[b!]   
\begin{center}
	\renewcommand{\arraystretch}{1.3}
\begin{tabularx}{\textwidth}{|@{}l|X|X@{}}
    \hline
    Method & Mean Absolute Error \\ 
    \hline 
    Multi-atlas propagation \cite{Ninon} & 131.4HU $\pm$ 60HU \\
    Context-Aware Generative Adversarial Network \cite{Nie} & 92.5HU $\pm$ 13.9HU \\
    Deep CNN \cite{XiaoHan} & 84.8HU $\pm$ 17.3HU \\
    Deep Boosted Regression & 68.6HU $\pm$ 15HU \\
    \hline
    \end{tabularx}
    \caption{Mean absolute error (MAE) in Houndsfield Units (HU) of state-of-the-art multi-atlas propagation method, two deep learning CT synthesis methods and proposed Deep Boosted Regression.}
    \label{MAE}
\end{center}
\end{table}

\section{Discussion and Conclusion}

In this work we proposed a new image-to-image translation network that is able to synthesise CT images from input MR images by gradually reducing the error using a separate boosting network. We validated the advantages of the recursive boosting model using a four-fold random bootstrapped validation with a 80:20 split that showed that the average difference between synthesised CT and ground-truth CT images was 68.6HU $\pm$ 15HU, compared to Burgos et al.'s method that achieved a MAE of 131.4HU $\pm$ 60HU. Other deep learning approaches reported a MAE of 92.5HU $\pm$ 13.9HU \cite{Nie} and 84.8HU $\pm$ 17.3HU \cite{XiaoHan}. However, while results are not directly comparable due to differing data, DBR reports state-of-the-art results on MAE among other deep learning approaches. 

To quantify the performance of the proposed Deep Boosted Regression method relative to the CT-based attenuation correction, the mean absolute percentage error (MAPE) within the head region only was used as the figure of merit. The obtained MAPE for the proposed method was 7.2\%, which showed an improvement to the state-of-the-art method \cite{Ninon}, which obtained MAPE of 14.3\%. As part of our future work, we will also investigate the impact of the synthesised CT images in radiotherapy treatment dose planning.

Furthermore, the success of the training highly depends on the registration quality of the MR/CT database. Even small inaccuracies in the registration can have a great influence on the training. An idea to circumvent the requirement of paired data is to incorporate a generative adversarial loss which provides a means of learning the context between CT and MR images from unpaired data. This has potential to provide a significant advantage in terms of the data availability for training due to the scarcity of accurately paired datasets, however, challenges in terms of validation emerge due to a missing ground truth. Moreover, we intend to extend the weighted patch sampling scheme to an adaptive sampling scheme that samples patches dynamically from areas with large residuals.

\section*{Acknowledgements}
This work was supported by an IMPACT studentship funded jointly by \linebreak Siemens and the EPSRC UCL Centre for Doctoral Training in Medical Imaging (EP/L016478/1). The research was also supported through the UK NIHR UCLH Biomedical Research Centre.


\bibliographystyle{splncs}
\bibliography{ReferencesTech}

\end{document}